\documentclass[12pt,epsfig]{article}

\usepackage{amssymb,amsmath}
\usepackage{epsfig,float}
\usepackage[hypertex,colorlinks=true,linkcolor=red,citecolor=blue]{hyperref}

\newcommand{\be}{\begin{eqnarray}}
\newcommand{\ee}{\end{eqnarray}}
\newcommand{\ba}{\begin{array}}
\newcommand{\ea}{\end{array}}

\newcommand{\bea}{\begin{eqnarray}}
\newcommand{\eea}{\end{eqnarray}}

\newcommand{\bi}{\begin{itemize}}
\newcommand{\ei}{\end{itemize}}

\begin{document}
\begin{titlepage}

\begin{flushright}
\begin{tabular}{l}
 CPHT-RR050.112016
\end{tabular}
\end{flushright}
\vspace{1.5cm}

\begin{center}
{\LARGE \bf
Backward
charmonium production
in $\pi N$ \\ collisions
}
\vspace{1cm}

\renewcommand{\thefootnote}{\alph{footnote}}

{\sc B.~Pire}${}^{1}$,
{\sc K.~Semenov-Tian-Shansky}${}^2$, %
{\sc L.~Szymanowski}${}^{3}$
\\[0.5cm]
\vspace*{0.1cm} ${}^1${\it
Centre de Physique Th\'eorique, {\'E}cole Polytechnique, CNRS, Universit\'e Paris-Saclay, F-91128 Palaiseau, France
                       } \\[0.2cm]
 \vspace*{0.1cm} ${}^2${\it
National Research Centre ``Kurchatov Institute'': Petersburg Nuclear Physics
Institute, RU-188300 Gatchina, Russia
                       } \\[0.2cm]
\vspace*{0.1cm} ${}^3$ {\it
 National Centre for Nuclear Research (NCBJ), PL-00-681 Warsaw, Poland
                       } \\[1.0cm]
{\it \large
\today
 }
\vskip2cm
{\bf Abstract:\\[10pt]} \parbox[t]{\textwidth}{
  The QCD collinear factorization framework allows one to describe exclusive backward production of a $J/\psi$ meson in pion-nucleon collisions in terms of pion-to-nucleon transition distribution amplitudes. We calculate the scattering amplitude at the leading order in the strong coupling constant and estimate the cross section of this reaction
in the backward kinematical region for a medium energy pion beam available at the J-Parc experimental facility.}
\vskip1cm
\end{center}

\vspace*{1cm}
\end{titlepage}

\setcounter{footnote}{0}

\section{Introduction}
\label{Sec_Intro}
\mbox

Besides leptoproduction experiments, hadronic facilities open
complementary accesses to the study of the partonic content of hadrons.
Indeed the collinear factorization theorems of quantum chromodynamics (QCD)
allow one to define universal hadronic matrix elements which enter scattering
amplitudes in both lepton-nucleon and meson-nucleon reactions in specific
kinematical conditions. This statement is true for both inclusive and exclusive
reactions. In the last two decades, we have witnessed  tremendous progress in the understanding of deeply virtual Compton scattering (DVCS) and deep meson
electroproduction within this framework. The detailed study of generalized parton
distributions (GPDs) \cite{GPD}, the relevant hadronic matrix elements, is a major goal
of modern hadronic physics.
Timelike processes such as the timelike Compton scattering (TCS)
\cite{TCS}
and exclusive Drell-Yan production in
$\pi N$ collisions
\cite{BDP} obey the same factorization properties and allow to
access the same GPDs. Exclusive charmonium production has
also been addressed in the same framework
\cite{Ivanov:2004vd}.

The extension of the collinear factorization approach to other processes
such as backward virtual Compton scattering and backward meson electroproduction, has been advocated
\cite{Frankfurt:1999fp,Pire:2004ie,Pire:2005ax}
although the corresponding factorization theorems have not yet been rigorously proven.
This leads to the definition of new hadronic matrix elements of three quark operators on the light cone,
the nucleon-to-meson transition distribution amplitudes (TDAs)
\cite{Pire:2016aqa}.
To motivate the validity of such a factorization regime, one requires the existence of
a large scale
$Q$,
which may be taken as the virtuality of the photon quantifying the electromagnetic
probe or the mass of a heavy quark in the case of heavy quarkonium production. This large
scale plays the role of the factorization scale and determines the magnitude of
the QCD coupling constant
$\alpha_s$.


The intense pion beam available at the Japan Proton Accelerator Research Complex (J-Parc)
(the pion beam momentum $P_\pi \sim 10 - 20$~GeV; center-of-mass energy squared
$W^2=m_N^2+m_\pi^2+2 E_\pi m_N \approx 2m_N P_\pi$)
opens the possibility to study hard
exclusive reactions such as lepton pair or charmonium production in
$\pi N$
collisions. This will provide new ways of testing the universality of GPDs
and TDAs. The recent feasibility study
\cite{JParc}
of forward lepton pair production suggests that GPDs can be accessed there.
Here we address the complementary case of backward charmonium production, the
perturbative QCD description of which involves pion-to-nucleon TDAs.
This process can be seen as the cross-channel counterpart of nucleon-antinucleon
annihilation into heavy quarkonium in association with a pion.
The description of this latter process within the collinear factorization approach
involving nucleon-to-pion TDAs was studied in Ref.~\cite{Pire:2013jva}.
The cross section estimates performed for the kinematical conditions of
\={P}ANDA@GSI-FAIR
lead to large enough production rates to be  experimentally accessed
\cite{PsiTDA,Singh:2016qjg}.

\section{Kinematics of the reaction}
\label{Sec_Kinematics}

\begin{figure}[H]
\begin{center}
\epsfig{figure= 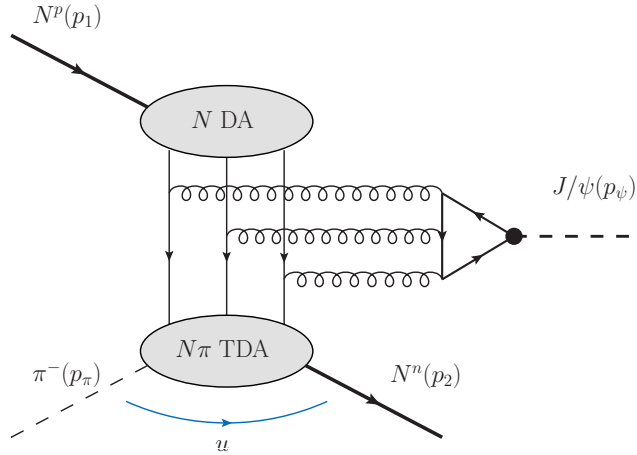 , height=6cm}
\end{center}
\caption{Collinear factorization of the
$\pi^-(p_\pi)+ N^p(p_1) \to N^n(p_2) + J/\psi(p_\psi)$
reaction in the $u$-channel regime.
DA stands for the distribution amplitude of the incoming nucleon;
$\pi \to N$
TDA stands for the transition distribution amplitude from a pion to a nucleon.}
\label{Fig_factorization}
\end{figure}

In the present paper we  consider the reaction
\be
\label{reacpsi}
\pi^-(p_{\pi}) \;+ N^p (p_1)  \; \to  J/\psi(p_{\psi}) + N^n (p_2) \,.
\label{reac}
\ee
The
$\pi N $
center-of-mass energy squared
$s=(p_\pi+p_1)^2 \equiv W^2$
and the charmonium mass squared
$M^2_\psi$
introduce the natural hard scale. In complete analogy with our
analysis of the nucleon-antinucleon annihilation process
\cite{Lansberg:2007se,Lansberg:2012ha}
we assume that this reaction admits a factorized description  in the
near-backward kinematical regime (see Fig.~\ref{Fig_factorization}), where
$|u| \equiv |\Delta^2|= |(p_2-p_\pi)^2| \ll W^2, \, M_\psi^2$.
This corresponds to the final nucleon moving almost in the direction of the initial  pion in
$\pi N$
center-of-mass system (CMS).

The
$z$-axis is chosen along the direction of the pion beam in the
meson-nucleon CMS frame.
We introduce the
light-cone vectors
$p, n$
satisfying
$ 2p \cdot n =1$.
The Sudakov decomposition of the relevant momenta  reads
\bea
&&
p_\pi = (1+\xi) p +\frac{m_\pi^2}{1+\xi}n\, ;
\nonumber \\ &&
p_1 =
\frac{2(1+\xi)m_N^2}{W^2+\Lambda(W^2,m_N^2,m_\pi^2)-m_N^2-m_\pi^2}\,
p +
\frac{W^2+\Lambda(W^2,m_N^2,m_\pi^2)-m_N^2-m_\pi^2}{2(1+\xi)} \,
n\,;   \nonumber \\ &&
\Delta \equiv (p_2-p_\pi)   =
-2\xi p + \left( \frac{ m_N^2-\Delta_T^2}{1-\xi } - \frac{m_\pi^2}{1+\xi}\right)n+\Delta_T;
\nonumber \\ &&
p_\psi =  p_1 - \Delta \, ; \ \ \ \
p_2= p_\pi +\Delta,
\label{Sudakov_decomposition}
\eea
where
\be
\Lambda(x,y,z)= \sqrt{x^2+y^2+z^2-2xy-2xz-2yz}
\label{Def_Mandelstam_f}
\ee
is the Mandelstam function and
$m_N$
and
$m_\pi$
stand respectively for the nucleon and pion masses.
The transverse direction in
(\ref{Sudakov_decomposition})
is defined with respect to the
$z$
direction and
$\xi$
is the  skewness variable
\be
\xi \equiv  -\frac{(p_2-p_\pi) \cdot n}{(p_2+p_\pi) \cdot n}.
\label{Def_xi}
\ee

Within the collinear factorization framework we neglect both the pion
and nucleon masses with respect to
$M_\psi$
and
$W$
and set
$\Delta_T=0$
within the coefficient function.
This results in the approximate expression for the  skewness variable
(\ref{Def_xi}):
\be
\xi \simeq \frac{M_\psi^2}{2 W^2-M_\psi^2}.
\label{Xi_collinear}
\ee
However, the approximation
(\ref{Xi_collinear})
can potentially affect the definition of the physical
domain of the reaction
(\ref{reac})
determined by the requirement
$\Delta_T^2 \le 0$,
where
\be
\Delta_T^2= \frac{1-\xi}{1+\xi} \left( u-2\xi \left[ \frac{m_\pi^2}{1+\xi}-
\frac{m_N^2}{1-\xi} \right] \right).
\label{Def_DT2}
\ee
To improve the approximate kinematical formulas in the vicinity of the threshold
it is sometimes convenient to keep partly the finite mass corrections resulting
in the modified expression for skewness variable
\be
\xi =\frac{M_\psi^2-m_N^2-u}{W^2 + \Lambda(W^2,m_N^2,m_\pi^2)+u-M_\psi^2-m_\pi^2}
+O(\frac{m_N^4}{W^4})+O(\frac{m_N^2 u}{W^4})+O(\frac{m_N^2 m_\pi^2}{W^4}).
\label{Xi_more_exact}
\ee

In order to control the validity of the kinematic approximations
(\ref{Xi_collinear}),
(\ref{Xi_more_exact})
it is instructive to consider the exact kinematics of the reaction
(\ref{reac})
in the
$\pi N$
CMS frame. In this frame the relevant momenta read:
\be
&&
p_\pi= \left(     \frac{W^2+m_\pi^2-m_N^2}{2W}    , \, \vec{p}_\pi\right);
\ \ \ \ \
p_\psi= \left(  \frac{W^2+M_\psi^2-m_N^2}{2W}, \, -\vec{p}_2 \right); \nonumber
\\ &&
p_1= \left( \frac{W^2+m_N^2-m_\pi^2}{2W} , \, -\vec{p}_\pi \right);
\ \ \ \ \
p_2= \left(  \frac{W^2+m_N^2-M_\psi^2}{2W}, \, \vec{p}_2 \right),
\ee
where
\be
|\vec{p}_\pi|= \frac{\Lambda(W^2,m_N^2,m_\pi^2)}{2W}; \ \ \
|\vec{p}_2|=\frac{\Lambda(W^2,m_N^2,M_\psi^2)}{2W}.
\ee
The CMS scattering angle
$\theta_u^*$
is defined as the angle between
$\vec{p}_\pi$
and
$\vec{p}_2$:
\be
\cos \theta_u^*= \frac{2W^2(u-m_N^2-m_\pi^2)+(W^2+m_\pi^2-m_N^2)(W^2+m_N^2-M_\psi^2)}
{\Lambda(W^2,m_N^2,m_\pi^2)\Lambda(W^2,m_N^2,M_\psi^2)}.
\label{CosTheta_exact}
\ee
The transverse momentum transfer squared
(\ref{Def_DT2})
is then given by
\be
\Delta_T^2=- \frac{\Lambda^2(W^2,M_\psi^2,m_N^2)}{4W^2}(1-\cos^2 \theta_u^*)
\label{DeltaT2_exact}
\ee
and the physical domain for the reaction
(\ref{reac})
is defined from the requirement that
$\Delta_T^2 \le 0 $.

\bi
\item
In particular, the backward kinematics regime
$\theta_u^*=0$
corresponds to
$\vec{p}_2$
along
$\vec{p}_\pi$,
which means that
$J/\psi$
is produced along
$-\vec{p}_\pi$ {\it i.e.}
in the backward direction. In this case
$u$
reaches its maximal value
\be
&&
u_{\max} \equiv \frac{2 \xi(m_\pi^2(\xi-1)+m_N^2(\xi+1))}{\xi^2-1}
\nonumber \\ &&
=m_N^2+m_\pi^2- \frac{(W^2+m_\pi^2-m_N^2)(W^2+m_N^2-M_\psi^2)}{2W^2}
+2|\vec{p}_\pi||\vec{p}_\psi|.
\label{Def_umax}
\ee
At the same time
$t=(p_2-p_1)^2$
reaches its minimal value
$t_{\min}$
($W^2+u_{\max}+t_{\min}=2m_N^2+m_\pi^2+M_\psi^2$).

Note that $u$ is negative, and therefore
$|u_{\max}|$
is the minimal possible absolute value of the momentum transfer squared.
It is for
$u \sim u_{\max}$
that one may expect to satisfy the requirement
$|u| \ll W^2, \, M_\psi^2$
which is crucial for the validity of the factorized description of
(\ref{reac})
in terms of
$\pi \to N$
TDAs and nucleon distribution amplitudes (DAs).

\item Another limiting value
 $\theta_u^*=\pi$
corresponds to
$\vec{p}_2$ along $-\vec{p}_\pi$
{\it i.e}
$J/\psi$
produced in the forward direction.
In this case $u$ reaches its minimal value
\be
u_{\min}=m_N^2+m_\pi^2- \frac{(W^2+m_\pi^2-m_N^2)(W^2+m_N^2-M_\psi^2)}{2W^2}
-2|\vec{p}_\pi||\vec{p}_\psi|.
\ee
At the same time
$t=(p_2-p_1)^2$
reaches its maximal value
$t_{\max}$.
The factorized description in terms of
$\pi \to N$
TDAs does not apply in this case as
$|u|$
turns to be of order of
$W^2$.
\ei

\section{Hard part of the $\pi^- + N^p \to  J/\psi + N^n $ amplitude}
\label{Sec_amplitude}
\mbox

The calculation of the
$\pi^- + N^p \to  J/\psi + N^n $
scattering amplitude follows the same main steps as the classical calculation of the
$J/\psi \to p + \bar{p}$ decay amplitude
\cite{Lepage:1980,Brodsky:1981kj,Chernyak:1983ej,Chernyak:1987nv}.
Assuming the factorization of small and large distance dynamics the hard part of the amplitude is computed within
perturbative QCD (pQCD). Large distance dynamics is encoded within the matrix elements of QCD light-cone operators between
the appropriate hadronic states.

The leading order amplitude of the
$J/\psi ~N^n$
production subprocess of
(\ref{reac})
is, up to the reverse of the direction of fermionic lines,
given by the sum of the same
three diagrams presented on Fig.~2 of
Ref.~\cite{Pire:2013jva}.

For the calculation of these diagrams, we apply the collinear approximation, neglecting both the
nucleon and pion masses and assuming
$\Delta_T=0$.
Therefore, the Sudakov decomposition
(\ref{Sudakov_decomposition})
reads as
\bea
&&
p_\pi \simeq (1+\xi) p \,; \ \ \ \
p_1 \simeq \frac{s}{(1+\xi)} n\,;   \ \ \ \ \
p_\psi \simeq 2 \xi p + \frac{s}{(1+\xi)} n \,; \ \ \ \ \Delta  \simeq -2 \xi p \,.
\eea
Also throughout our calculation we set
$
M_\psi  \simeq  2m_c \equiv \bar{M}
$
with $\bar{M} = 3\, {\rm GeV}$.


Below, we summarize our conventions for the relevant light-cone matrix elements encoding the
soft dynamics :
\bi
\item
The leading twist-$3$
$uud$ $\pi^-$-to-neutron
($\pi^- \to n$)
TDAs are defined from the Fourier transform
\be
{\cal F} \equiv
(p \cdot n)^3 \int
\left[
\prod_{j=1}^3 \frac{d \lambda_j}{2 \pi}
\right]
e^{i \sum_{k=1}^3 x_k \lambda_k (p \cdot n)}
\label{Fourier}
\ee
of the $n\pi^-$ matrix element of the
trilinear antiquark operator on the light cone
\be
M^{(\pi^- \to n)}_{\rho \tau \chi}(\lambda_1 n, \lambda_2 n ,\lambda_3 n)=  \langle n(p_2)| \varepsilon_{c_1 c_2 c_3}
\bar{u}_{\rho}^{c_1}(\lambda_1n)
\bar{u}_{\tau}^{c_2}(\lambda_2n)
\bar{d}_{\chi}^{c_3}(\lambda_3n)
|\pi^-(p_\pi) \rangle.
\ee
Namely,
\be
&&
4  {\cal F} M^{(\pi^- \to n)}_{\rho \tau \chi}(\lambda_1 n, \lambda_2 n ,\lambda_3 n)
\nonumber \\ &&
= \delta(x_1+x_2+x_3-2 \xi) i \frac{f_N}{f_\pi}
\sum_{\rm Dirac \atop structures} s^{(\pi \to N)}_{\rho \tau, \, \chi} H_s^{
(\pi^- \to n)}(x_1,x_2,x_3, \xi, \Delta^2),
\label{FT_defining_npi_TDAs}
\ee
where the sum goes over the eight
leading twist Dirac structures
$s^{(\pi \to N)}_{\rho \tau, \, \chi}$
built of
$p$, $n$, $\Delta_T$,
charge conjugation matrix
$C$
and the Dirac spinor
$\bar{U}(p_2)$.
The explicit form of these Dirac structures is worked out in the
Appendix~\ref{App_A},
which contains also the relation
of the parametrization of the leading twist
$ \pi^- \to n$
TDAs to the conventional
$n \to \pi^-$
TDAs introduced in
Ref.~\cite{Lansberg:2007ec}.

\item For the leading twist antinucleon DAs, we employ the standard
parametrization of Ref.~\cite{Chernyak:1983ej} involving
three invariant functions
$V^p$,
$A^p$
and
$T^p$
(see also Appendix~B of Ref.~\cite{Lansberg:2012ha}).

\item For the light-cone wave function of
$J/\psi$
heavy quarkonium ,we use the so-called nonrelativistic
wave function suggested in Ref.~\cite{Chernyak:1983ej}.
\ei

The leading order and leading twist amplitude of the reaction
(\ref{reac})
admits the following parametrization,
\footnote{We  adopt Dirac's ``hat'' notation $\hat{v} \equiv v_\mu \gamma^\mu$.}
\bea
&&
{\cal M}^{s_1 s_2}_\lambda= {\cal C} \frac{1}{{\bar M}^5 } \Big[
\bar{U}(p_2,s_2 )\hat{\cal E}^*(\lambda) \gamma_5 U(p_1, s_1) {\cal I}(\xi,\Delta^2)
\nonumber \\ &&
-\frac{1}{m_N}  \bar{U}(p_2,s_2 )\hat{\cal E}^*(\lambda)
\hat{\Delta}_T \gamma_5 U(p_1, s_1) {\cal I}'(\xi,\Delta^2)
\Big],
\label{Amplitude_master}
\eea
where $\cal E$ is the charmonium polarization vector and
$\bar{U}$, $U$
stand  for the nucleon Dirac spinors.

The calculation of the
three
Born order diagrams
yields the same result for the integral
convolutions
$\{{\cal I}, \, {\cal I'} \} (\xi,\Delta^2)$
as for
$J/\psi \; \pi$
production in
$\bar{p}p$
annihilation up to the obvious replacement
of nucleon-to-pion ($N \to \pi$) TDAs
with pion-to-nucleon ($\pi \to N$) TDAs
introduced in (\ref{FT_defining_npi_TDAs}).
The explicit expressions for
${\cal I}$ and ${\cal I'}$
can be found in Eqs.~(13) and (15)
of Ref.~\cite{Pire:2013jva}.
The overall numerical factor
$\cal C$
in
(\ref{Amplitude_master})
is expressed as:
\bea
{\cal C}= (4 \pi \alpha_s)^3 \frac{f_N^2 f_\psi}{f_\pi}  \, \frac{10}{81},
\eea
where
$\alpha_s$
stands for the strong coupling,
$f_\pi=93$~MeV
is the pion weak decay constant,
$f_\psi$
determines the normalization
of the wave function of heavy quarkonium and
$f_N$
determines the value of the nucleon wave function at the origin.
The normalization constant
$f_\psi$
is extracted from the charmonium leptonic decay width
$\Gamma(J/\psi \to e^+  e^- )$.
With the values quoted in
Ref.~\cite{PDG2016},
we get
$|f_\psi|= 416 \pm 5 \; {\rm MeV}$.

\section{Estimates of the cross section}

To work out the cross section formula, we
square the amplitude
(\ref{Amplitude_master})
and average (sum) over spins of initial (final) nucleons.
Staying at the leading twist accuracy, we account for the
production of transversely  polarized
$J/\psi$. Summing over the transverse polarizations we find
\bea
|\overline{\mathcal{M}_{T}}|^2 =
\sum_{\lambda_T} \left( \frac{1}{2} \sum_{s_1 s_2}
{\cal M}^{s_1 s_2}_{\lambda_T} ({\cal M}^{s_1 s_2}_{\lambda_T})^* \right).
\eea

The leading twist differential cross section of
$\pi +N   \to J/\psi + N$
then reads
\bea
&&
\frac{d \sigma}{d \Delta^2}= \frac{1}{16 \pi \Lambda^2(s,m_N^2,m_\pi^2) } |\overline{\mathcal{M}_{T}}|^2 \nonumber  \\ &&
=\frac{1}{16 \pi \Lambda^2(s,m_N^2,m_\pi^2)}
\frac{1}{2} |\mathcal{C}|^2 \frac{2(1+\xi)}{\xi {\bar{M}}^8}  \left( |\mathcal{I}(\xi, \Delta^2)|^2 - \frac{\Delta_T^2}{m_N^2} |\mathcal{I}'(\xi, \Delta^2)|^2 \right),
\label{CS_def_delta2}
\eea
where
$\Lambda(x,y,z)$
is defined in
(\ref{Def_Mandelstam_f}).

In order to get a rough estimate of the cross section we use the simple nucleon exchange model for
$\pi \to N$ TDAs
suggested in
Ref.~\cite{Pire:2011xv}.
We do not expect that the inclusion of the spectral part for
$\pi \to N$
TDAs
\cite{Pire:2010if,Lansberg:2011aa}
would be essential to draw a conclusion on  the feasibility of the
relevant experiment. The refinement of the present description will be done
in course of availability of precise experimental data.

The predictions of the cross-channel nucleon exchange model of
Ref.~\cite{Pire:2011xv}
for
$n \to \pi^-$
TDAs within the parametrization
(\ref{Old_param_TDAs})
are summarized in Eqs. (25)--(27) of Ref.~\cite{Pire:2013jva}.
Employing the results of the Appendix~\ref{App_A},
we conclude that $\pi^- \to  n$ TDAs within this model are
expressed as
\bea
&&
\big\{ V_{1,2}, \, A_{1,2} , \, T_{1,2,3}  \big\}^{(\pi^- \to  n)} ( 
x_{1,2,3}, \xi,\Delta^2)\Big|_{N(940)} \nonumber \\ &&
=  \sqrt{2} \big\{V_{1,2}, \, A_{1,2} , \, T_{1,2,3}    \big\}^{(p  \to \pi^0 )} ( 
-x_{1,2,3}, -\xi,\Delta^2)\Big|_{N(940)};
\nonumber \\ &&
T_4^{(\pi^- \to n)} ( 
x_{1,2,3}, \xi,\Delta^2)\Big|_{N(940)}=0.
\label{Nucleon_exchange_contr_VAT_pi_minus_crossed}
\eea
Note that nucleon-to-pion TDAs within
the cross-channel nucleon exchange model have purely
the Efremov-Radyushkin-Brodsky-Lepage (ERBL)-like support.
As the result the convolution integrals
$\cal I$, $\cal I'$
within this model turn to be real since
the poles in the corresponding integrands
are located either on the crossover trajectories
which separate the Dokshitzer-Gribov-Lipatov-Altarelli-Parisi (DGLAP)-like
and the ERBL-like support regions of
$\pi \to N$ TDAs%
\footnote{For the definition of the ERBL-like and DGLAP-like
support regions of TDAs, see
Ref.~\cite{Pire:2010if}.},
or within the DGLAP-like support regions.

The convolution integrals
$\cal I$, $\cal I'$
within the simple nucleon pole model
(\ref{Nucleon_exchange_contr_VAT_pi_minus_crossed})
are expressed as
\bea
&&
{\cal I}(\xi, \Delta^2)\Big|_{N(940)}=-\sqrt2
\frac{  f_\pi \,   g_{\pi NN}  m_N (1+\xi) } {   (\Delta^2-m_N^2) (1-\xi )} M_0;
\nonumber \\ &&
{\cal I}'(\xi, \Delta^2)\Big|_{N(940)}=- \sqrt2
\frac{  f_\pi \,   g_{\pi NN}  m_N   } {    (\Delta^2-m_N^2)  } M_0,
\eea
where
$M_0$
is given by Eq.~(19) of Ref.~ \cite{Pire:2013jva}.
Note that the integral convolution
$M_0$
also occurs in the well-known expression for the
$J/\psi \to \bar{p} p$
decay width within the pQCD approach
\cite{Chernyak:1987nv}:
\be
\Gamma_{J/\psi \to p \bar{p}}
= (\pi \alpha_s)^6 \frac{1280 f_\psi^2 f_N^4 }{243 \pi {\bar{M}^9}} |M_0|^2.
\label{Charm_dec_width}
\ee

As the phenomenological input for the cross section estimate we may employ
different solutions for the leading twist nucleon DAs $V^p$, $A^p$, $T^p$.
Similarly to the
case of the charmonium decay width, our result  depends strongly
both on the form of the input leading twist nucleon DAs and the value of
$\alpha_s$.

Analogously to
Ref.~\cite{Pire:2013jva}, we
have chosen to  present our results for the
$\pi^- p \to J/\psi \,  n $
cross section with the value of
$\alpha_s$
fixed by the requirement that the given
phenomenological solution reproduces the experimental
$J/\psi \to N \bar{N}$
decay width from the pQCD expression
of Ref.~\cite{Chernyak:1987nv}.
The corresponding values of
$\alpha_s$
for several phenomenological solutions for the leading twist nucleon DA
are summarized in the Table~1 of
Ref.~\cite{Pire:2013jva}.

Phenomenological solutions for nucleon DAs that are
largely concentrated in the end point regions such as
the Chernyak-Ogloblin-Zhitnitsky
\cite{Chernyak:1987nv}
or
King-Sachrajda 
\cite{King:1986wi}
require smaller values of
$\alpha_s \sim 0.25$
to reproduce the experimental value of
$\Gamma_{J/\psi \to p \bar{p}}$.
The solutions  which are close to the
asymptotic form of the nucleon DA
$\phi^N_{\rm as}(y_{1,2,3})\equiv
V^p_{\rm as}(y_{1,2,3})-A^p_{\rm as}(y_{1,2,3})=120 y_1 y_2 y_3$
like the Bolz-Kroll 
\cite{Bolz:1996sw}
and Braun-Lenz-Wittmann next-to-leading order
 model \cite{Braun:2006hz}
require rather large values of
$\alpha_s \sim 0.4$
to reproduce the experimental value of
$\Gamma_{J/\psi \to p \bar{p}}$
with Eq.~(\ref{Charm_dec_width}).
We refer the reader to
Refs.~\cite{Stefanis_DrNauk,Brambilla:2004wf}
for the discussion and critics of the available phenomenological
models of the leading twist nucleon DAs.

Note that with the aforementioned assumptions the predictions of
the cross-channel nucleon exchange model for
$\pi \to N$
TDAs
(\ref{Nucleon_exchange_contr_VAT_pi_minus_crossed})
for the cross section
(\ref{CS_def_delta2})
can be seen as being independent of the particular form
of the phenomenological solution for the leading twist nucleon DAs.
The cross section can be be expressed through the experimental value 
\cite{PDG2016}
of 
the decay width 
$\Gamma_{J/\psi \to p \bar{p}}=0.197 \pm 0.009$~KeV:
\be
\frac{d \sigma}{d \Delta^2}
\Big|_{N(940)}
= \frac{40}{27} \frac{g_{\pi NN}^2 \Gamma_{J/\psi \to p \bar{p}} \, \bar{M} m_N^2}{(m_N^2-\Delta^2)^2 \Lambda^2(W^2, m_N^2, m_\pi^2)}
\frac{1+\xi}{\xi}
\left[
\frac{(1+\xi)^2}{(1-\xi)^2}- \frac{\Delta_T^2}{m_N^2}
\right].
\label{CS_DA_indep}
\ee
Let us stress that we employ this model only as the first very
rough cross section estimate intended for the preliminary feasibility
studies at J-Parc kinematical conditions.
Once the experimental data would be available, its description
within the suggested factorization mechanism would require both
the detailed discussion on the form of the leading twist nucleon DAs and
$\pi \to N$ TDAs
as well as the estimation of the possible
higher twist and $\alpha_s$ corrections.

\begin{figure}[H]
 \begin{center}
  \epsfig{figure= 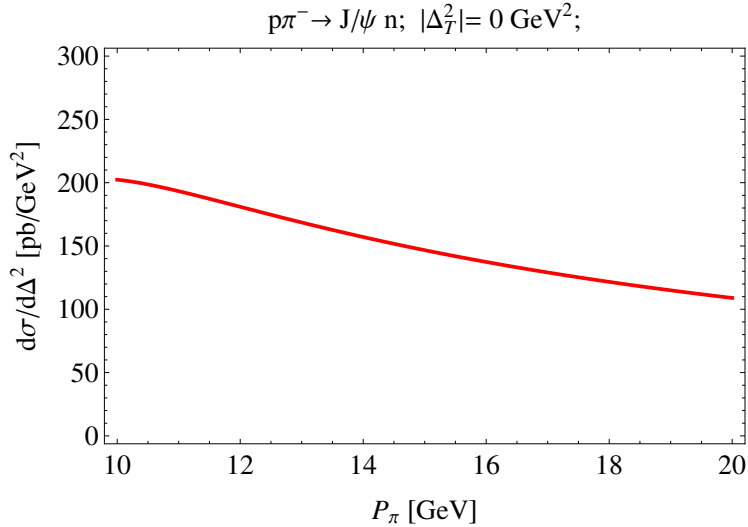 , height=7.0cm}
    \end{center}
     \caption{Differential cross section
$\frac{d \sigma}{d \Delta^2}$ for  $\pi^- p \to J/\psi \,  n$
as a function of
the pion beam momentum $P_\pi$ ($W^2 \approx 2m_N P_\pi$)
for
$\Delta_T^2=0$.
Our estimate is based on the nucleon exchange contribution to the 
$\pi \to N$ TDAs  and is independent of the specific form of the 
leading twist nucleon DA [see Eq.~(\ref{CS_DA_indep}) and the corresponding discussion].
}
\label{Fig_CS_W2}
\end{figure}

On Fig.~\ref{Fig_CS_W2},
we show our estimates of the differential cross section
$\frac{d \sigma}{d \Delta^2}$ for $\pi^- p \to J/\psi \,  n$
(\ref{CS_def_delta2})
as a function of
the pion beam momentum
$P_\pi$
for the exactly  backward charmonium production
($\Delta_T^2=0 \Leftrightarrow \theta^*_\pi=0$).
The range of the pion beam momentum
$10\,{\rm GeV} \le P_\pi \le 20\,{\rm GeV}$
corresponds to the J-Parc experimental setup.

\begin{figure}[H]
 \begin{center}
\epsfig{figure= 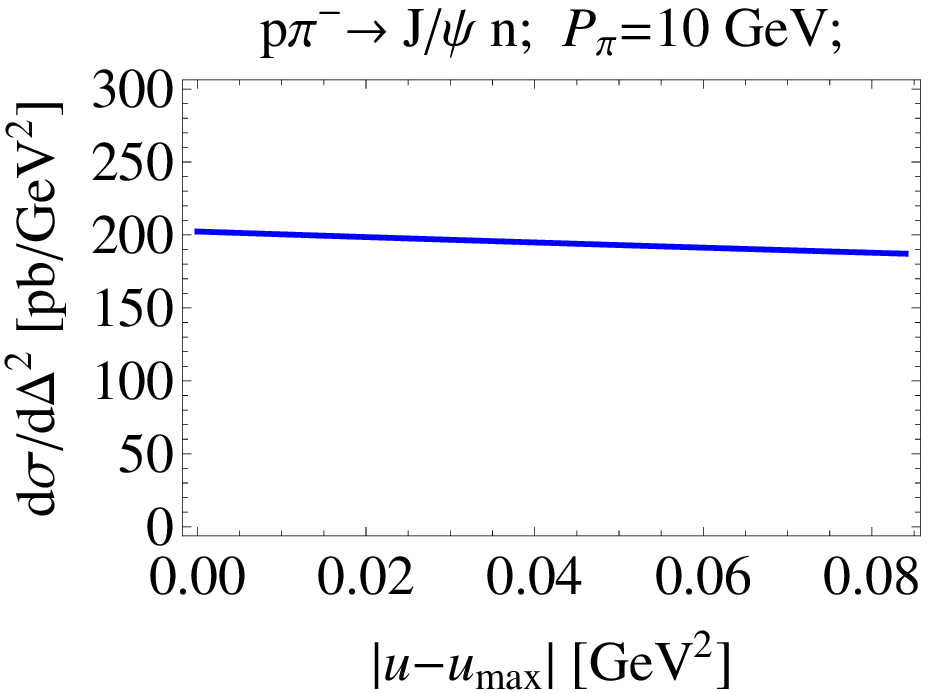, height=5.0cm}
 \epsfig{figure= 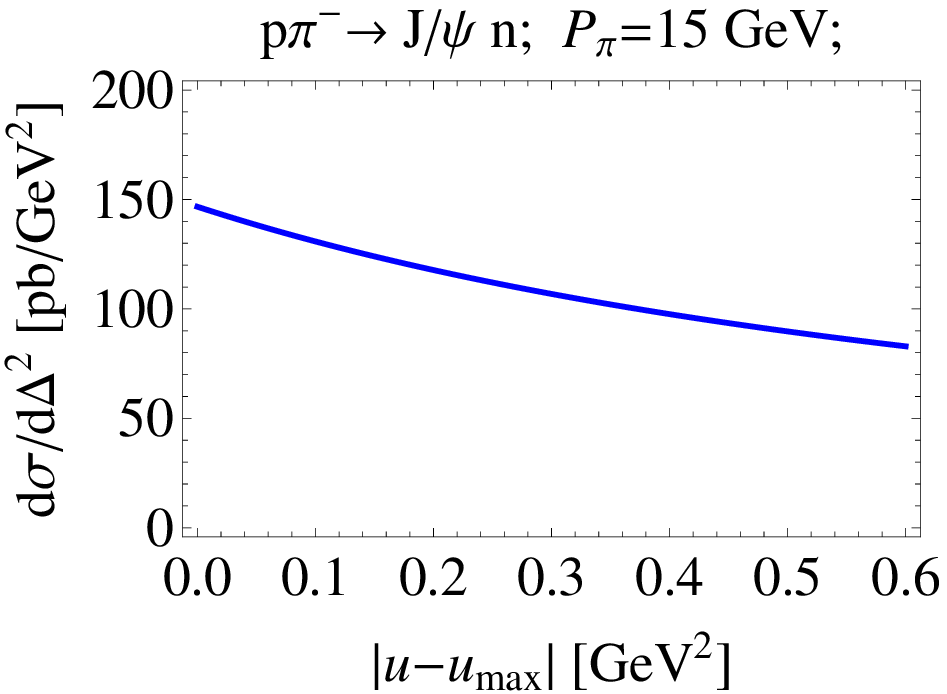, height=5.0cm}
 \epsfig{figure= 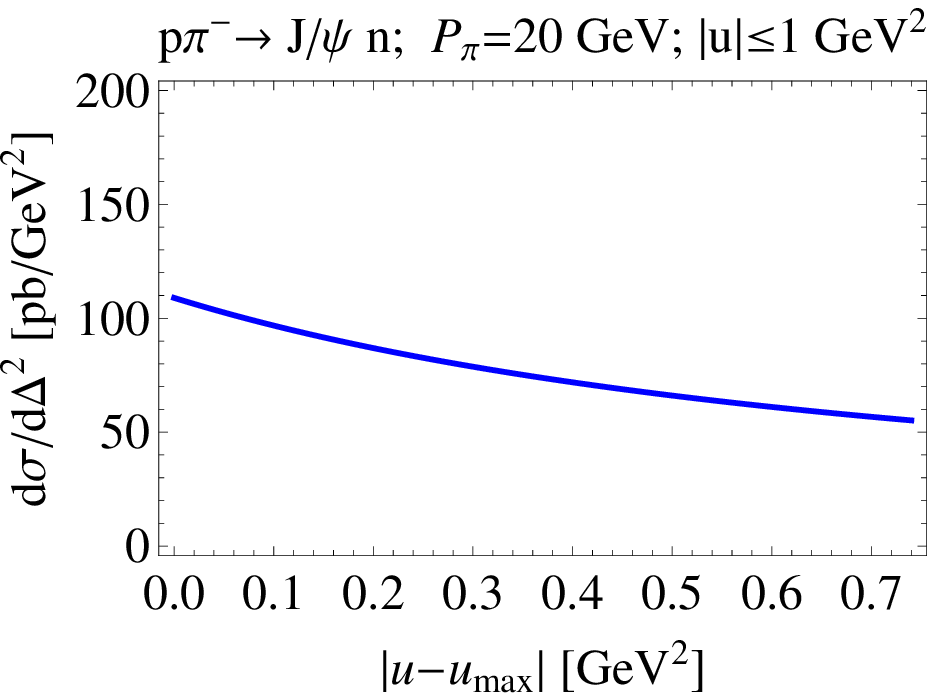, height=5.0cm}
   \end{center}
     \caption{ Differential cross section
$\frac{d \sigma}{d \Delta^2}$ for  $\pi^- p \to J/\psi \,  n$
as a function of
$|u-u_{\max}|$
for
$P_\pi=10, \, 15, \, 20$~GeV for $|u| \le 1$~GeV$^2$.}
\label{Fig_CS_u_minus_umax}
\end{figure}

On Fig.~\ref{Fig_CS_u_minus_umax},
we show the differential cross section
$\frac{d \sigma}{d \Delta^2}$
for
 $\pi^- p \to J/\psi \,  n$
as a function of
$|u-u_{\max}|$ for $|u| \le 1$~GeV$^2$,
where
$u_{\max}$
is the threshold value
(\ref{Def_umax})
of the momentum transfer squared.
We show the result for several values of the pion beam energy
$P_\pi$.
In order to better describe the cross section behavior for
$\Delta_T^2 \ne 0$
we employ the  exact value
(\ref{Def_umax})
for the maximal cross-channel momentum transfer squared.

\begin{figure}[H]
 \begin{center}
\epsfig{figure= 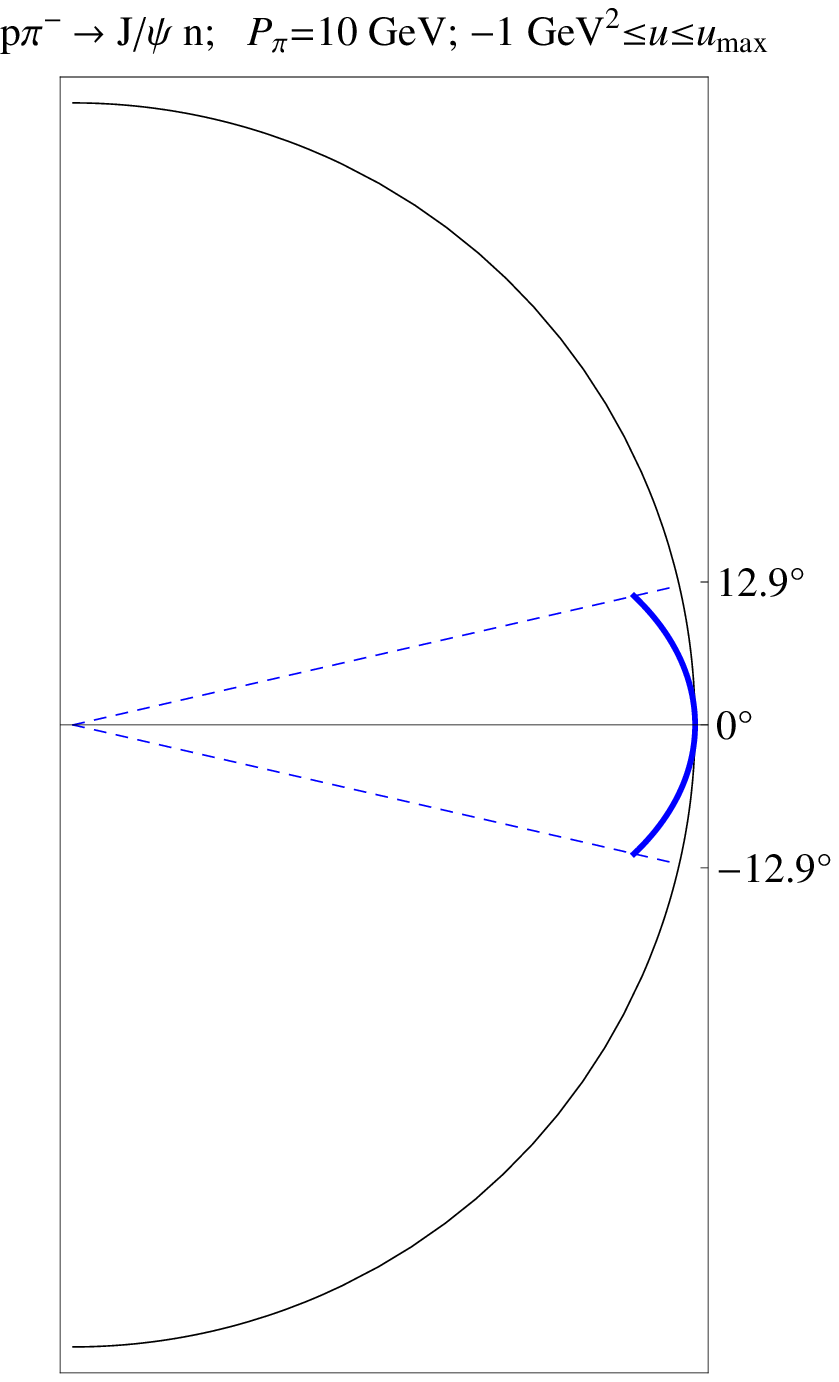, height=8.0cm}
 \epsfig{figure= 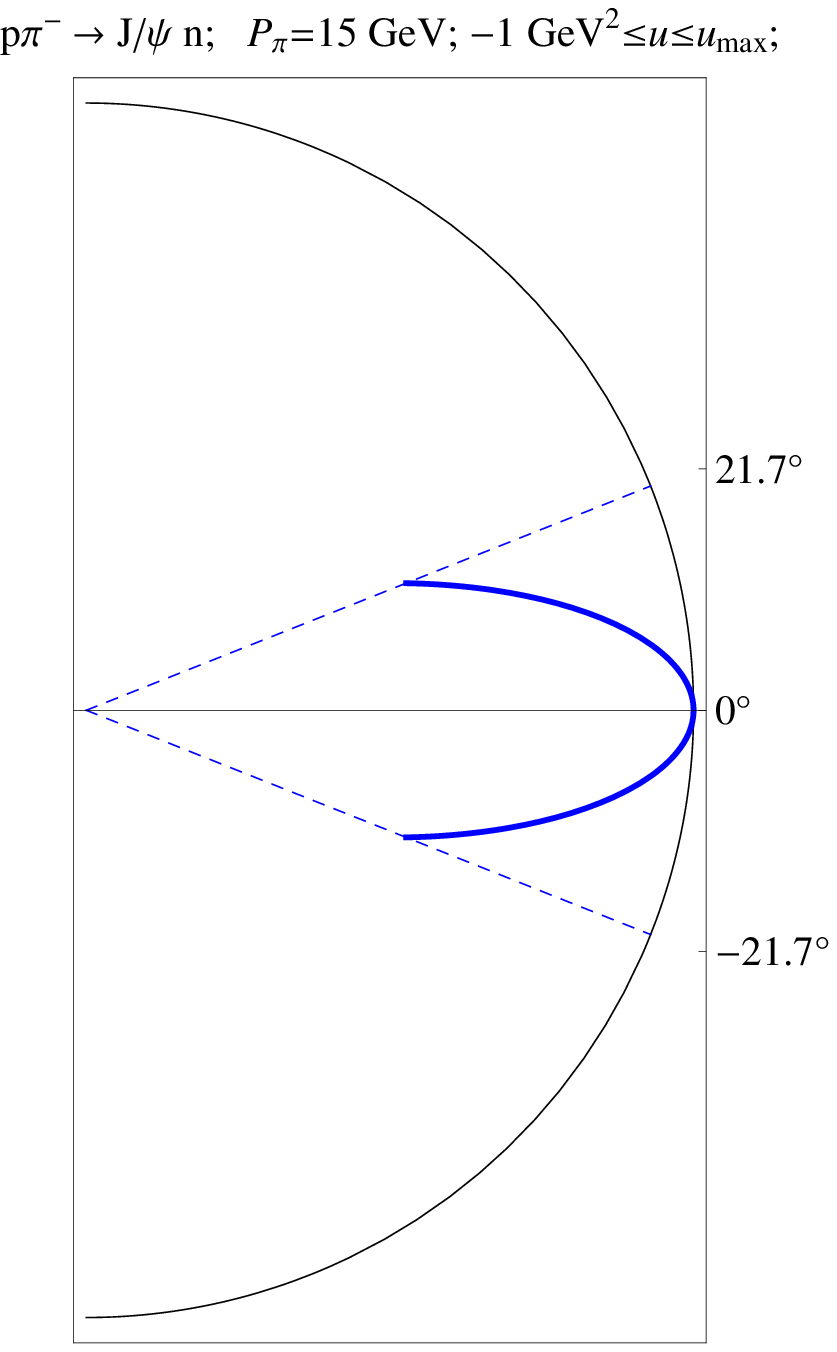, height=8.0cm}
 \epsfig{figure= 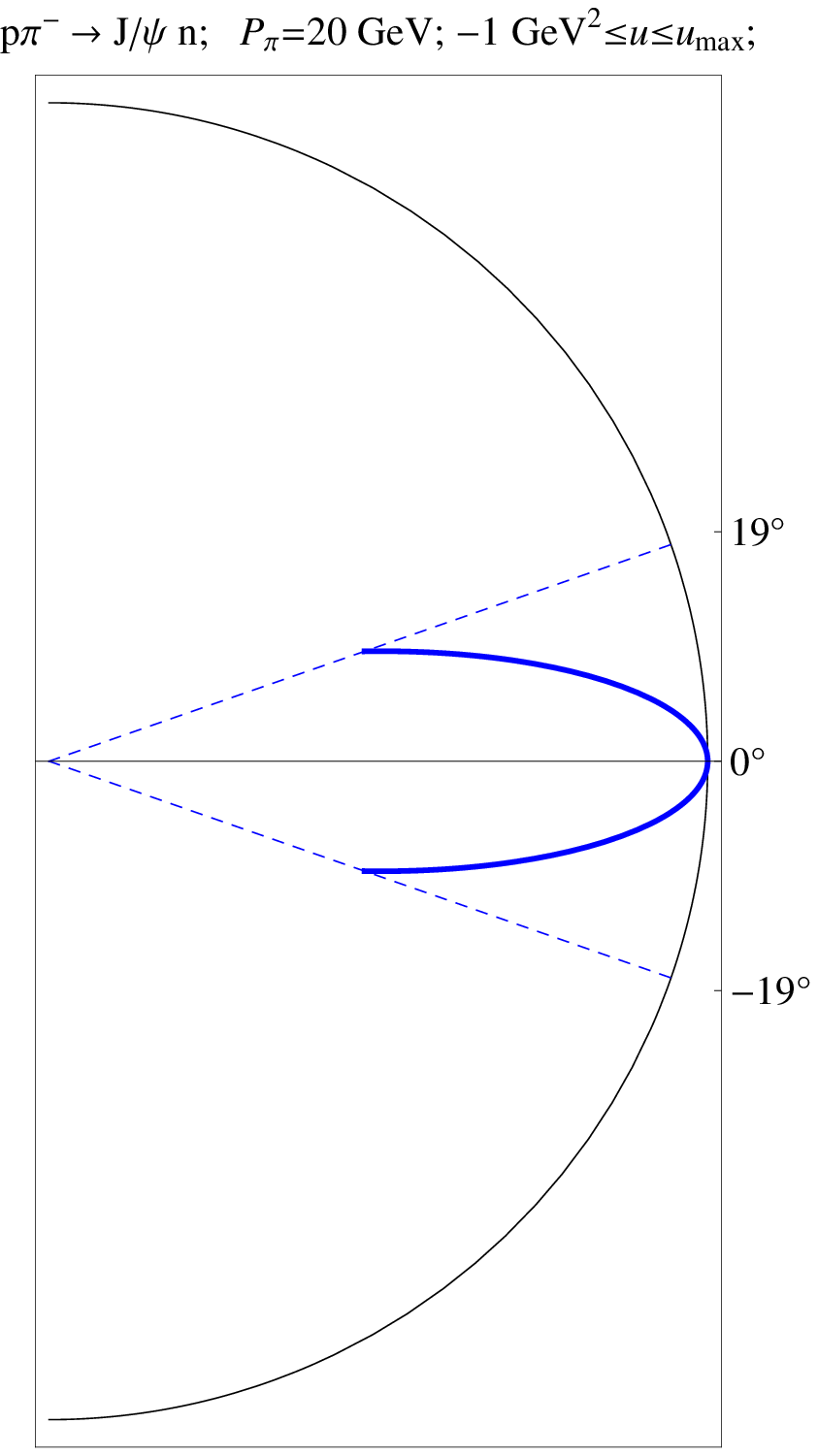, height=8.0cm}
   \end{center}
     \caption{ Angular distribution for the
$d \sigma/d \Delta^2$
cross section for the near-backward
charmonium production
($\theta_\pi^*\simeq 0$)
for   $-1 \, {\rm GeV}^2 \le  \Delta^2 \le u_{\max}$.
Dashed lines show the effect of the cutoff
$|\Delta^2| \le 1$  GeV$^2$
for the values of the pion CMS scattering angle $\theta_\pi^*$.  }
\label{Fig_CS_theta}
\end{figure}

On Fig.~\ref{Fig_CS_theta}
we show the characteristic center-of-mass angular distribution for the
$d \sigma/d \Delta^2$
cross section for the backward factorization regime visualized on the polar plot with the
polar angle being the pion CMS scattering angle
$\theta_\pi^*$
[see Eq.~(\ref{CosTheta_exact}) for the definition].
We present the ratio
\be
\frac{ \frac{d \sigma} {d \Delta^2} (W^2, \theta_\pi^*)}
{\frac{d \sigma} {d \Delta^2} (W^2, \theta_\pi^*=0)}
\ee
as a function of
$\theta_\pi^*$
showing the result for $P_\pi=10, \, 15, \, 20$~GeV and for
$-1 \, {\rm GeV}^2 \le  \Delta^2 \le u_{\max}$,
where
$u_{\max}$
is the threshold value
(\ref{Def_umax})
of the momentum transfer squared.
With the dashed lines we show the effect of the cutoff
$|\Delta^2| \le 1$  GeV$^2$
for the values of the CMS scattering angle
$\theta_\pi^*$
[see the discussion around Eq.~(\ref{CosTheta_exact})].

Since these rates are certainly within the experimental reach of the
J-Parc experiment, the study of the reaction
(\ref{reac})
will provide a valuable universality test for the TDA approach since the same
TDAs also arise in the description of
$N \bar{N} \to \gamma^* \pi$
\cite{Lansberg:2012ha},
$N \bar{N} \to J/\psi \pi$
\cite{PsiTDA,Singh:2016qjg}
and backward pion electroproduction off a nucleon
$\gamma^* N \to \pi N$
\cite{Lansberg:2011aa}.

\section{Conclusions}

In this paper we address the reaction
$\pi^- +\, N^p \to J/\psi \,+ N^n$
which may be studied in the  J-Parc facility. We argue that
this reaction may be analyzed within
the pQCD framework. It will not only help to quantitatively disentangle resonance production from the universal
hadronic background but also will   provide valuable information on the hadronic structure encoded in pion-to-nucleon
TDAs. Pion-to-nucleon TDAs supply complementary information
with respect to partonic distributions diagonal in quantum numbers such as
common parton distributions and GPDs.

Within the kinematical range accessible at  J-Parc, we provide the predictions
for the $\pi^- +\, N^p \to J/\psi \,+ N^n$ cross section
using a simple nucleon pole model for
$\pi \to N$
TDAs. The obtained  cross section estimates give hope of experimental
accessibility of the reaction. A specific feasibility study
similar to that recently performed for accessing
$N \to \pi$
TDAs
at \={P}ANDA
\cite{Singh:2014pfv,PsiTDA,Singh:2016qjg}
should be performed with J-Parc experimental efficiencies, as it has
been done for exclusive forward lepton pair production at J-Parc
\cite{JParc}.

It is  worth mentioning that the mass of the charm quark may not be large enough
for our leading order (in
$\alpha_s$)
and leading twist analysis to be sufficient to
describe the data. More work is certainly needed to go beyond the Born approximation
for the hard amplitude, in particular because the timelike nature of the hard probe
is often accompanied by large
$O(\alpha_s)$
corrections
\cite{Muller:2012yq}.

\section*{Acknowledgements}
\mbox
We acknowledge useful discussions with Professor Shunzo Kumano, Wen-Chen Chang,
Jen-Chieh Peng and Shinya Sawada.

This work is partly supported by Grant No.
 2015/17/B/ST2/01838 by the National Science Center in Poland, by the French grant
 ANR PARTONS (Grant No. ANR-12-MONU-0008-01), by the COPIN-IN2P3 agreement, by the
 Labex P2IO and by the Polish-French collaboration agreement Polonium.
K.S. acknowledges the support from the Russian Science Foundation (Grant No. 14-22-00281).

\setcounter{section}{0}
\setcounter{equation}{0}
\renewcommand{\thesection}{\Alph{section}}
\renewcommand{\theequation}{\thesection\arabic{equation}}

\section{Crossing $\pi \to N$ TDAs to $N \to \pi$ TDAs}
\label{App_A}
\mbox

We have the parametrization for the nucleon-to-pion
($N \to \pi$)
TDAs defined through the Fourier transform of the
$\pi N$
matrix element of the trilinear quark operator on the light cone.
The parametrization involves
eight
invariant functions, each being the function of three
longitudinal momentum fractions, the skewness variable,
the momentum transfer squared as well as of the factorization scale.

Throughout this paper, we make use of the parametrization of
Ref.~\cite{Lansberg:2007ec},
where only three invariant functions turn out to be relevant in the
$\Delta_T=0$
limit.
Let us consider the neutron-to-$\pi^-$ $uud$ TDA,
\bea
&&
4 (p \cdot n)^3 \int \left[ \prod_{j=1}^3 \frac{d \lambda_j}{2 \pi}\right]
e^{i \sum_{k=1}^3 \tilde{x}_k \lambda_k (p \cdot n)}
 \langle     \pi^-(p_\pi)|\,  \varepsilon_{c_1 c_2 c_3} u^{c_1}_{\rho}(\lambda_1 n)
u^{c_2}_{\tau}(\lambda_2 n)d^{c_3}_{\chi}(\lambda_3 n)
\,|n(p_N,s_N) \rangle
\nonumber \\ &&
= \delta(\tilde{x}_1+\tilde{x}_2+\tilde{x}_3-2 \tilde{\xi}) i \frac{f_N}{f_\pi}\Big[  V^{(n \to \pi^-)}_{1}(\tilde{x}_{1,2,3}, \tilde{\xi} ,\tilde{\Delta}^2)  (  \hat{p} C)_{\rho \tau}(U^+)_{\chi}
\nonumber \\ &&
+A^{(n \to \pi^-)}_{1}(\tilde{x}_{1,2,3}, \tilde{\xi},\tilde{\Delta}^2)  (  \hat{p} \gamma^5 C)_{\rho \tau}(\gamma^5 U^+ )_{\chi}
 +T^{(n \to \pi^-)}_{1}(\tilde{x}_{1,2,3}, \tilde{\xi},\tilde{\Delta}^2)  (\sigma_{p\mu} C)_{\rho \tau }(\gamma^\mu U^+ )_{\chi}
 \nonumber \\ &&
 + m_N^{-1} V^{(n \to \pi^-)}_{2} (\tilde{x}_{1,2,3}, \tilde{\xi},\tilde{\Delta}^2)
 ( \hat{ p}  C)_{\rho \tau}( \hat{\tilde{\Delta}}_T U^+)_{\chi}
+ m_N^{-1}
 A^{(n \to \pi^-)}_{2}(\tilde{x}_{1,2,3}, \tilde{\xi},\tilde{\Delta}^2)  ( \hat{ p}  \gamma^5 C)_{\rho\tau}(\gamma^5  \hat{\tilde{\Delta}}_T  U^+)_{\chi}
  \nonumber \\ &&
+ m_N^{-1} T^{(n \to \pi^-)}_{2} (\tilde{x}_{1,2,3}, \tilde{\xi} ,\tilde{\Delta}^2) ( \sigma_{p \tilde{\Delta}_T} C)_{\rho \tau} (U^+)_{\chi}
+  m_N ^{-1}T^{(n \to \pi^-)}_{3} (\tilde{x}_{1,2,3}, \tilde{\xi},\tilde{\Delta}^2) ( \sigma_{p\mu} C)_{\rho \tau} (\sigma^{\mu \tilde{\Delta}_T}
 U^+)_{\chi}
 \nonumber \\ &&
+ m_N^{-2} T^{(n \to \pi^-)}_{4} (\tilde{x}_{1,2,3}, \tilde{\xi},\tilde{\Delta}^2)  (\sigma_{p \tilde{\Delta}_T} C)_{\rho \tau}
(\hat{ \tilde{\Delta}}_T U^+)_{\chi} \Big]
 \nonumber \\ &&
\equiv
\delta(\tilde{x}_1+\tilde{x}_2+\tilde{x}_3-2 \tilde{\xi}) i \frac{f_N}{f_\pi}
\sum_{\rm Dirac \atop structures} s^{( N \to \pi)}_{\rho \tau, \, \chi} H_s^{(n \to \pi^-)}( \tilde{x}_1,\tilde{x}_2,\tilde{x}_3, \tilde{\xi}, \tilde{\Delta}^2).
 \label{Old_param_TDAs}
\eea
We adopt Dirac's ``hat'' notation $\hat{v} \equiv v_\mu \gamma^\mu$;
$\sigma^{\mu\nu}= \frac{1}{2} [\gamma^\mu, \gamma^\nu]$; $\sigma^{v \mu} \equiv v_\lambda \sigma^{\lambda \mu}$;
$C$
is the charge conjugation matrix and
$U^+= \hat{p} \hat{n} \, U(p_N,s_N)$
is the large component of the nucleon spinor.
$f_\pi=93$ MeV
is the pion weak decay constant and
$f_N$
determines the value of the nucleon wave function at the origin.

Note that the
$N \to \pi$
TDA
(\ref{Old_param_TDAs})
is defined with respect to its natural kinematical variables.
Namely the cross-channel momentum transfer is
$\tilde{\Delta}=p_\pi-p_N$,
and the skewness parameter
$\tilde{\xi}$
is defined from the longitudinal momentum transfer
between the pion and nucleon,
$$
\tilde{\xi}
 \equiv  -\frac{( p_\pi-p_N) \cdot n}{(p_\pi+p_N) \cdot n}
$$
[{\it i.e.} it differs by the sign from the definition
(\ref{Def_xi})
natural for the reaction
(\ref{reac})].

\begin{figure}[H]
 \begin{center}
\epsfig{figure= 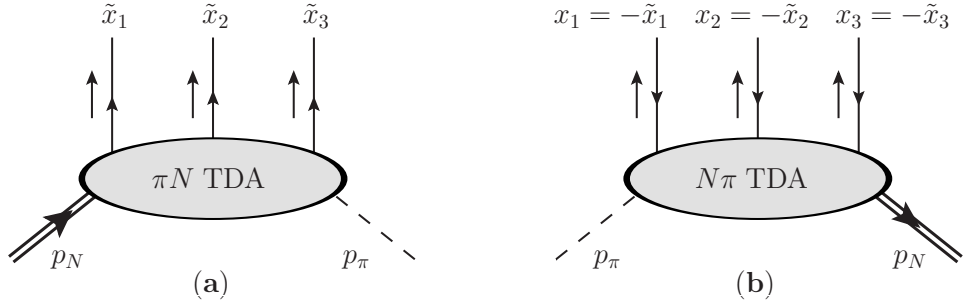, height=4.0cm}
   \end{center}
     \caption{Small arrows show the direction of the longitudinal momentum flow in the ERBL-like
      regime for the following:
{\bf (a)} The longitudinal momentum flow for
$N \to \pi$ TDAs
defined in
(\ref{Old_param_TDAs}).
The longitudinal momentum transfer is $(p_\pi-p_N) \cdot n \equiv \tilde{\Delta} \cdot n$.
{\bf (b)}: The longitudinal momentum flow for
$\pi \to N$ TDAs
defined in
(\ref{FT_defining_npi_TDAs}).
The longitudinal momentum transfer is $(p_N-p_\pi) \cdot n \equiv  \Delta \cdot n$.
Arrows on the nucleon and quark (antiquark) lines show the direction of flow of the baryonic charge.}
\label{Fig_Flow}
\end{figure}


Now, we would like to express
pion-to-nucleon
($\pi \to N$)
TDAs through
($N \to \pi$)
TDAs occurring in
(\ref{Old_param_TDAs}).
For this issue, we apply the Dirac conjugation
(complex conjugation and convolution with
$\gamma_0$ matrices
in the appropriate spinor indices)
for both sides of Eq.~(\ref{Old_param_TDAs}) and
compare the result to the definition of  $\pi \to N$ TDAs
(\ref{FT_defining_npi_TDAs}):
\be
&&
-4 (p \cdot n)^3 \int \left[ \prod_{j=1}^3 \frac{d \lambda_j}{2 \pi}\right]
e^{-i \sum_{k=1}^3 \tilde{x}_k \lambda_k (p \cdot n)}
 \langle    n(p_N,s_N) |\,  \varepsilon_{c_1 c_2 c_3} \bar{u}^{c_1}_{\rho}(\lambda_1 n)
\bar{u}^{c_2}_{\tau}(\lambda_2 n) \bar{d}^{c_3}_{\chi}(\lambda_3 n)
\,| \pi^-(p_\pi) \rangle
\nonumber \\ &&
=
-\delta(\tilde{x}_1+\tilde{x}_2+\tilde{x}_3-2 \tilde{\xi}) i \frac{f_N}{f_\pi}
\sum_s   \underbrace{(\gamma_0^T)_{\tau \tau'} \left[ s_{\rho' \tau', \chi'}^{(N \to \pi)} \right]^\dag
(\gamma_0)_{\rho'\rho }
(\gamma_0)_{\chi' \chi}}_{s_{\rho \tau, \chi}^{(\pi \to N)}} H_s^{(N \to \pi)}
(\tilde{x}_{1},\tilde{x}_{2},\tilde{x}_{3}, \tilde{\xi} ,\Delta^2).
\nonumber \\ &&
\label{DiracConjTDA}
\ee

For the relevant Dirac structures we get
\be
&&
(v_1^{(\pi \to N)})_{\rho \tau, \chi}= (C \hat{p})_{\rho \tau} \bar{U}^+_\chi;
\nonumber \\ &&
(a_1^{(\pi \to N)})_{\rho \tau, \chi}= (C \hat{p} \gamma_5)_{\rho \tau} \left(\bar{U}^+ \gamma_5 \right)_\chi;
\nonumber \\ &&
(t_1^{(\pi \to N)})_{\rho \tau, \chi}= -(C \sigma_{p \mu})_{\rho \tau} \left(\bar{U}^+ \gamma_\mu \right)_\chi;
\nonumber \\ &&
(v_2^{(\pi \to N)})_{\rho \tau, \chi}= (C \hat{p})_{\rho \tau}  \left( \hat{\tilde{\Delta}}_T\bar{U}^+ \right)_\chi=-
 (C \hat{p})_{\rho \tau}  \left( \hat{ \Delta}_T\bar{U}^+ \right)_\chi;
\nonumber \\ &&
(a_2^{(\pi \to N)})_{\rho \tau, \chi}= (C \hat{p} \gamma_5)_{\rho \tau} \left( \bar{U}^+ \hat{\tilde{\Delta}}_T \gamma_5 \right)_\chi=-
 (C \hat{p} \gamma_5)_{\rho \tau} \left( \bar{U}^+  \hat{\Delta}_T \gamma_5 \right)_\chi
\nonumber \\ &&
(t_2^{\pi \to N)})_{\rho \tau, \chi}= -(C \sigma_{p \tilde{\Delta}_T})_{\rho \tau} \left(\bar{U}^+ \right)_\chi=(C \sigma_{p  \Delta_T})_{\rho \tau} \left(\bar{U}^+ \right)_\chi;
\nonumber \\ &&
(t_3^{(\pi \to N)})_{\rho \tau, \chi}= (C \sigma_{p \mu})_{\rho \tau}
\left(\bar{U}^+ \sigma_{  \mu \tilde{\Delta}_T} \right)_\chi=
- (C \sigma_{p \mu})_{\rho \tau}
\left(\bar{U}^+ \sigma_{  \mu  \Delta_T} \right)_\chi;
\nonumber \\ &&
(t_4^{(\pi \to N)})_{\rho \tau, \chi}= -(C \sigma_{p \tilde{\Delta}_T})_{\rho \tau}
\left(\bar{U}^+  \hat{\tilde{\Delta}}_T\right)_\chi=
-(C \sigma_{p \Delta_T})_{\rho \tau}
\left(\bar{U}^+  \hat{\Delta}_T \right)_\chi\,,
\ee
where we switch to the definition of momentum transfer
natural for the reaction (\ref{reac}):
$\tilde{\Delta} \to -\Delta$.
$\bar{U}^+\equiv \bar{U}(p_N) \hat{n} \hat{p}$
stands for the large component of the
$\bar{U}(p_N)$
Dirac spinor.

The flow of the longitudinal momentum for
$N \to \pi$
TDAs defined as in Eq.~(\ref{Old_param_TDAs}) and
$\pi \to N$
TDAs defined as in Eq.~(\ref{FT_defining_npi_TDAs}) is presented on Fig.~\ref{Fig_Flow}.
By switching to the variables
$\xi=-\tilde{\xi}$
and
$x_i=-\tilde{x}_i$
natural for the reaction
(\ref{reac})
and
$\tilde{\Delta}^2 \to \Delta^2$
and comparing
(\ref{DiracConjTDA})
to
(\ref{FT_defining_npi_TDAs})
we conclude that
\be
&&
\left\{
V_{1,\, 2}^{(\pi^- \to n)}, \,
A_{1,\, 2}^{(\pi^- \to n)}, \,
T_{1,\, 2, \, 3, \,4}^{(\pi^- \to n)}
\right\}
(x_{1,2,3}, \xi, \Delta^2)
\nonumber \\ &&
=
\left\{
V_{1,\, 2}^{ (n \to \pi^- )}, \,
A_{1,\, 2}^{  (n \to \pi^-)}, \,
T_{1,\, 2, \, 3, \,4}^{  (n \to \pi^-)}
\right\}
(-x_{1,2,3}, -\xi, \Delta^2).
\ee

\end{document}